\RequirePackage{ifpdf}
\ifpdf % We are running pdfTeX in pdf mode
\documentclass[pdftex]{sigma}
\else
\documentclass{sigma}
\fi

\numberwithin{equation}{section}

\newcommand\Om\Omega

\newcommand\minus\backslash

\def\1{\'{\i}}

\def\>#1{{\mathbf#1}}

\begin{document}

\allowdisplaybreaks

\renewcommand{\thefootnote}{$\star$}

\renewcommand{\PaperNumber}{097}

\FirstPageHeading

\ShortArticleName{Quantum Exactly Solvable Radial Systems on Curved Space}

\ArticleName{A Family of Exactly Solvable Radial\\ Quantum  Systems on Space of Non-Constant\\ Curvature with Accidental Degeneracy\\ in the Spectrum\footnote{This
paper is a contribution to the Proceedings of the Workshop ``Supersymmetric Quantum Mechanics and Spectral Design'' (July 18--30, 2010, Benasque, Spain). The full collection is available at \href{http://www.emis.de/journals/SIGMA/SUSYQM2010.html}{http://www.emis.de/journals/SIGMA/SUSYQM2010.html}}}

\Author{Orlando RAGNISCO and Danilo RIGLIONI}

\AuthorNameForHeading{O.~Ragnisco and D.~Riglioni}

\Address{Dipartimento di Fisica Universit\'a Roma Tre and Istituto Nazionale di Fisica Nucleare,\\ Sezione di Roma, I-00146 Roma, Italy}
\Email{\href{mailto:ragnisco@fis.uniroma3.it}{ragnisco@fis.uniroma3.it}, \href{mailto:riglioni@fis.uniroma3.it}{riglioni@fis.uniroma3.it}}
\URLaddress{\url{http://webusers.fis.uniroma3/~ragnisco/}}

\ArticleDates{Received October 05, 2010, in f\/inal form December 07, 2010;  Published online December 15, 2010}

\Abstract{A novel family of exactly solvable quantum systems  on curved space is presented. The family is the quantum version of the classical Perlick family, which comprises all maxi\-mal\-ly superintegrable 3-dimensional Hamiltonian systems  with spherical symmetry.  The high number of symmetries (both geometrical  and dynamical) exhibited by the classical systems   has a counterpart  in the  accidental degeneracy in the spectrum  of  the quantum systems. This family of quantum problem is completely solved with the techniques of the SUSYQM (supersymmetric quantum mechanics). We also analyze in detail the ordering problem ari\-sing in the quantization of the kinetic term of the classical Hamiltonian, stressing the link existing between two physically meaningful quantizations: the geometrical quantization and the position dependent mass quantization.}


\Classification{(81S10; 81R12; 31C12)}

\section{Introduction}

A Hamiltonian system of $N$ degrees of freedom is said to be completely integrable, in the Arnold--Liouville  sense, if it possesses $N$ functionally independent globally def\/ined
and single valued  integrals of motion in involution~\cite{arnold}.  In analogy to the classical mechanics, a quantum mechanical system
with $N$ degrees of freedom is called integrable if it is equipped with a set of $N-1$ algebraically independent observables (self-adjoint operators densely def\/ined on a suitable Hilbert space) all of them commuting
with the Hamiltonian itself. The maximum number of independent observables commuting with the Hamiltonian is $2N-2$; in that case out of this set one can extract at least two (complete) sets of $N$ commuting observables, both containing the Hamiltonian.  Such a system is called maximally superintegrable (M.S.).

From a classical point of view these systems are characterized by the so called ``closed orbit property'' for bounded motions, and from a quantum point of view these systems present a~maximal degeneracy in the discrete spectrum.
Over the years the M.S.\ systems have been extremely useful models in the description of physical phenomena, as for instance the well known oscillator and Kepler radial problems.
Recently,  exactly solvable 1-dimensional systems have been proposed, above all in quantum mechanics, as models describing electronic properties in condensed matter physics, for instance in semiconductors and quantum dots  \cite{gritsev, quesneA, quesneB, cruz, mustafa, tanaka};
the main feature of these systems is the presence of an ef\/fective mass in the Schr\"odinger equation, that can be considered  as a position-dependent  particle mass.
Since, from a classical  point of view,  there exists a complete classif\/ication of all radial M.S.\ systems (f\/lat case Bertrand theorem~\cite{bertrand}, curved space Perlick theorem~\cite{perlick}), the main goal of this paper is to ``translate''
these classical systems in the quantum context. Speaking about quantum M.S.\ systems many authors have recently stressed the existing relations of such quantum systems with systems studied in supersymmetric quantum mechanics (SUSYQM)~\cite{marquette} and in fact in this paper we will achieve our goal by using the powerful tools of the (SUSYQM) in particular the intertwining techniques that in the end allow us  to obtain the spectrum and the eigenfunctions in a pure algebraic way.

\section{Perlick's family of classical M.S.\ systems}\label{section2}

Let us consider the standard radial Hamiltonian describing a particle subject to a central force:
\begin{gather*}
 H(P_r, P_\theta, P_\phi, r , \theta ,\phi)= \frac{1}{2}P_r^2 + \frac{L^2}{2r^2} + V(r).
\end{gather*}
A theorem proved by Bertrand in the nineteenth century \cite{bertrand} establishes that the only radial potentials such that all bounded orbits are closed turn out to be:
\begin{gather*}
V(r) = \omega^2r^2 , \qquad V(r) = -\frac{\mu}{r}.
\end{gather*}
We can obtain a wider class of such systems if we take in exam the more general Hamiltonian:
\begin{gather}\label{genham}
H(P_r, P_\theta, P_\phi, r , \theta, \phi) = T   + V(r) = \frac{f(r)}{2}P_r^2 + \frac{L^2}{2r^2} +V(r).
\end{gather}
The kinetic term of this new class of Hamiltonian systems can be regarded as that of a particle moving according to a geodesic motion in a complete Riemannian manifold $\Re^3$ with the metric:
\begin{gather*}
ds^2 = \frac{dr^2}{f(r)} + r^2 \big(d\theta^2 + \sin^2(\theta) d\phi^2\big).
\end{gather*}
Considering this new scenario, Perlick managed to f\/ind a classif\/ication of the pair of functions $(f(r) ; V(r))$ admitting stable closed orbit property~\cite{perlick}.
This classif\/ication is indeed the most general~\cite{nostro} and it contains as particular cases other systems with the closed orbit property as for example the so called multifold Kepler systems and Darboux III system~\cite{ikatayama}.
The Perlick's systems can be classif\/ied in two parametric families that can be viewed as a deformation of the two original Bertrand systems:

\medskip

\noindent
Family I (Kepler type)
\begin{gather}\label{perl1}
ds^2 = \frac{dr^2}{\beta^2 (1+k^2r^2)} + r^2\big(d\theta^2 +\sin^2(\theta)d\phi^2\big), \qquad V(r) =- \mu \sqrt{\frac{1}{r} + k^2}+G.
\end{gather}
Family II (oscillator type)
\begin{gather*}
ds^2 = \frac{2\big(1-Dr^2 \pm \sqrt{(1-Dr^2)^2-k^2r^4 }\big)}{\beta^2( (1-Dr^2)^2-k^2r^4)}dr^2 + r^2\big(d\theta^2 +\sin^2(\theta)d\phi^2\big),
\\
V(r) =  G \mp \frac{r^2}{ 1-Dr^2 \pm \sqrt{(1-Dr^2)^2-k^2r^4}}.
\end{gather*}
 In the above formulas, $G$ is just an additive constant, and indeed it could be harmlessly removed. However we prefer to keep it in order to be consistent with the formalism
used in our previous papers \cite{enciso,nostro}.
These systems were proven to be M.S.\ in a rigorous way in 2009 \cite{enciso} and def\/ine a complete
classif\/ication of all radial problems in a Riemannian manifold with radial symmetry.

\section{The generalized Kepler problem}\label{section3}

As said before, the Hamiltonians with radial symmetry turn out to be of physical  interest above all in the quantum context. So,  this section will be devoted to give a short introduction of  the quantization problem of systems def\/ined in a non f\/lat space and to provide the tools needed to solve them exactly.
The main problem is how to def\/ine the quantum analog of the kinetic term in~\eqref{genham}, since there is an order ambiguity in its quantization $T(r,p) \rightarrow \hat{T}(r,p)$.
This problem always arises when in the kinetic energy term there is a dependence on the coordinates \cite{vonros,quesneA,mustafa}, which comes from the nonzero curvature of the space.
So, our Hamiltonian $H= f P_r^2 + V(r)$ could be quantized for example as:
\begin{gather}\label{vonr}
-f^a \partial_r f^{1-a-b} \partial_r f^b + \frac{l(l+1)}{r^2} + V(r), \qquad  a,b \in {\mathbb{R}}.
\end{gather}
Since our goal is to preserve the M.S.\ property of the classical Perlick's systems  at the quantum level, we will  set aside, for a moment, the choice of the parameters $a$, $b$ and start  directly from a~M.S.\ quantum system.
To this aim we choose the Perlick~I space with the parameter $\beta=1$; this system is universally known as the generalized Kepler problem, namely the Kepler problem on a~space of constant curvature \cite{schrodinger,infield,higgs,leemon,pogosian,carinena}.
As many other M.S.\ quantum systems~\cite{marquette}, this system is indeed linked to a so-called shape-invariant system.

\subsection{Shape-invariant systems and intertwining technique}\label{section3.1}

Let us introduce the 1-dimensional  hyperbolic Kepler Hamiltonian, namely the radial part of the 3-dimensional Schr\"odinger equation on a space of constant negative curvature (``hyperbolic space''~\cite{infield}):
\begin{gather*}
\hat{H}_q = -\partial_r^2 + \frac{k^2 q(q-1)}{\sinh^2(kr)} - 2\mu k \frac{\cosh(kr)}{\sinh(kr)}.
\end{gather*}
It should be noted that the parameter $k$ is linked to the scalar curvature $R$ through the relation $R = -6k^2$ and in fact in the limit $k \rightarrow 0$ we recover the radial part of the usual f\/lat Kepler problem.
This Hamiltonian system is a member  of a small, though extremely relevant,  family of systems called shape-invariant systems: the reason for this name will be  clear soon.

These systems can be solved easily by the  ``intertwining technique''~\cite{samsonov}, which  can be seen as a generalization of the Dirac ladder-operator technique.
To clarify this point, let us def\/ine the  ``prepotential'' function:
\[
W_q(r) = -\frac{\mu}{q}r + q\ln(\sinh(kr)), \qquad q^2<\frac{\mu}{k}, \qquad q>0
\]
and the following two ladder operators:
\begin{gather*}
A_q^{\dagger} = \partial_r + {W'}_q(r) = \partial_r -\frac{\mu}{q} + qk\coth(kr),
\\
-A_q = \partial_r - {W'}_q(r)  =  \partial_r +\frac{\mu}{q} - qk\coth(kr).
\end{gather*}
The Hamiltonian turns out to be factorized by the ladder operators:
\begin{gather*}
A^{\dagger}_q A_q = \hat{H}_q -\epsilon_q , \qquad \epsilon_q = -\frac{\mu^2}{q^2} - k^2q^2 .
\end{gather*}
Since $A_q e^{W_q(r)} = 0 $ we have:{\samepage
\begin{gather*}
A_q^\dagger A_q e^{W_q(r)} = (\hat{H}_q - \epsilon_q)e^{W_q(r)} = 0 \quad \Rightarrow \quad \hat{H}_q e^{W_q(r)} = \epsilon_q e^{W_q(r)},
\end{gather*}
therefore we can read $e^{W_q(r)} = \psi_q (r) $ as an eigenfunctions of the Hamiltonian $\hat{H}_q$.}

 The crucial property turns up by intertwining the two ladder operators:
\begin{gather*}
A_q A^{\dagger}_q = H_{q+1} -\epsilon_q,
\end{gather*}
which shows that we get  the same Hamiltonian system,  apart from  the value of the parame\-ter~$q$ that turns out to be changed to  $q+1$. This is indeed the exceptional property of the shape invariant systems.
 This property allows us to f\/ind in a purely algebraic way the spectrum and the eigenvectors.
In fact, let us def\/ine the following relations:
\begin{gather*}
A^{\dagger}_{q+1} A_{q+1} \psi_{q+1}(r) = ( H_{q+1} - \epsilon_{q+1}) \psi_{q+1}(r)\\
\qquad{}
  =(A_q A^{\dagger}_q + \epsilon_q -\epsilon_{q+1}) \psi_{q+1} = 0.
\end{gather*}
Multiplying by $A^{\dagger}$ from the  left:
\begin{gather*}
(A^{\dagger}_q A_q + \epsilon_q -\epsilon_{q+1}) (A^{\dagger}_q\psi_{q+1}) = 0
\end{gather*}
and f\/inally:
\begin{gather*}
\hat{H}_{q} (A^{\dagger}_q \psi_{q+1}) = \epsilon_{q+1}(A^{\dagger}_q \psi_{q+1})
\end{gather*}
therefore $\psi_{1,q}(r) = A^{\dagger}_q \psi_{q+1}$ can be considered the f\/irst excited state of $H_q$.

 By iterating this procedure it is possible to gain the whole set of bound states eigenfunctions  as well as  their eigenvalues. In particular the eigenvalues read:
\begin{gather*}
\hat{H}_q \psi_{n,q}(r) = \epsilon_{q+n} \psi_{n,q}(r) = -\left(\frac{\mu^2}{(q+n)^2}+k^2(q+n)
^2\right)\psi_{n,q}(r), \qquad  n \in {\mathbb{N}},
\end{gather*}
while, up to a normalization factor,  the {\it  finite} set of the eigenfunctions will be:
\begin{gather*}
\psi_{n,q}(r) = \prod_{i=0}^{n-1} A^{\dagger}_{q+i} \psi_{0,q+n}(r),\qquad n\leq N_{\max},  \qquad (q+N_{\max})^2 = \left[\frac{\mu}{k}\right], \qquad N_{\max} \in \mathbb{N}.
\end{gather*}
As stated at the beginning of this section this system can be seen as the $k$ deformation of the radial part of the hydrogen atom Hamiltonian: in fact,  in the limit $k \rightarrow 0$, its eigenvalues (and eigenfunctions) turn out to be exactly those of the hydrogen atom:
\[
\epsilon_s = -\frac{\mu^2}{s^2}, \qquad s = q+n.
\]

\subsection{Hyperbolic Kepler and Perlick systems}\label{section3.2}

The radial part of the hyperbolic Kepler system is an exactly solvable 1-dimensional quantum system. In this subsection we are going to see how to link this 1-dimensional problem to one of the 3-dimensional problems associated to the metric~\eqref{perl1}.
This link can be found easily by making the following point transformation:
\begin{gather*}
r' = \frac{\sinh (kr)}{k}
\end{gather*}
yielding{\samepage
\begin{gather*}
\hat{H}_q = -\frac{1}{2} \partial_r^2 + \frac{k^2 q(q-1)}{2 \sinh^2(kr)} - \mu k \coth(kr) \\
\qquad {}\Rightarrow \quad -\frac{1}{2}\big(1+k^2{r'}^2\big) \partial_{r'}^2- \frac{k^2r'}{2}\partial_{r'} + \frac{q(q-1)}{2{r'}^2} -\mu \sqrt{\frac{1}{{r'}^2}+k^2},
\end{gather*}
namely the quantization~\eqref{vonr} with parameters $a=\frac{1}{2}$, $b=0$.}

The scalar product will change to the new one:
\begin{gather*}
\langle \psi_{n,q}|\psi_{n',q'}\rangle  =  \int_0^\infty \psi_{n,q}^*(r) \psi_{n',q'}(r) dr \quad \Rightarrow \quad \int_0^\infty \psi_{n,q}^*(r') \psi_{n',q'}(r') \frac{dr'}{\sqrt{1+k^2{r'}^2}}.
\end{gather*}
Finally, in order to get a geometrically coherent Hamiltonian system, we perform a similarity transformation that (of course) doesn't change the spectrum, but modif\/ies the expressions of the eigenfunctions
\begin{gather*}
\epsilon_{n,q} = \int_0^\infty \frac{{r'}^2 dr}{\sqrt{1+k^2{r'}^2}} \frac{\psi_{n,q}^*(r')}{r'} \left(\frac{1}{r'}\hat{H}r'\right) \frac{\psi_{n,q}(r')}{r'}.
\end{gather*}
We def\/ine this new operator as the  Laplace--Beltrami (LB) Hamiltonian:
\begin{gather*}
\hat{H}_{\rm LB}=\frac{1}{r'} \hat{H} r' = - \frac{1}{2}\left(\big(1+k^2{r'}^2\big)\partial_{r'}^2 + \frac{2}{r'}\partial_{r'} + 3k^2{r'}\partial_{r'} \right) + \frac{q(q-1)}{2{r'}^2} - \mu\sqrt{\frac{1}{{r'}^2} +k^2} -\frac{k^2}{2}
\end{gather*}
with eigenfunctions:
\begin{gather*}
\tilde{\psi}_{n,q}(r') = \frac{1}{r'} \prod_{i=0}^{n-1} A^{\dagger}_{q+i}(r') e^{W_{q+n}(r')}
\end{gather*}
orthogonal to each other with the new weight function $w(r')$ in the semiline $r' \in [0,\infty)$:
\begin{gather*}
w(r')dr' = \frac{{r'}^2 dr'}{\sqrt{1+k^2{r'}^2}}.
\end{gather*}
The problem solved above is still a 1-dimensional problem on the semiline. Then in order to establish an exact correspondence
between this 1-dimensional problem and the generalized Kepler problem we have to  ``upgrade'' this 1-dimensional problem  to a 3-dimensional one.
We def\/ine the standard 3-dimensional angular momentum operator in angular variables:
\begin{gather*}
\hat{L}^2 =-\left( \partial_\theta^2 +\cot(\theta) \partial_\theta +  \frac{1}{\sin^2(\theta)}\partial_\phi^2\right)
\end{gather*}
and their spherical harmonic eigenfunctions:
\begin{gather*}
\hat{L}^2 Y_{l,m}(\theta,\phi) = l(l+1) Y_{l,m}(\theta, \phi).
\end{gather*}
It is possible now write down the 3-dimensional Hamiltonian operator:
\begin{gather*}
\hat{H}_{\rm LB3d} = - \frac{1}{2}\left(\big(1+k^2{r'}^2\big)\partial_{r'}^2 + \frac{2}{r'}\partial_{r'} + 3k^2{r'}\partial_{r'} \right) + \frac{\hat{L}^2}{2{r'}^2} - \mu\sqrt{\frac{1}{{r'}^2} +k^2} -\frac{k^2}{2},
\\
\hat{H}_{\rm LB3d} \tilde{\psi}_{n,l+1}(r')Y_{l,m}(\theta,\phi) = \left(-\frac{\mu^2}{2(n+l+1)^2} - \frac{k^2(n+l+1)^2}{2}\right) \tilde{\psi}_{n,l+1}(r')Y_{l,m}(\theta,\phi).
\end{gather*}
We can recognize this Hamiltonian to be exactly the quantum version of the 3-dimensional gene\-ra\-lized Kepler system in which the kinetic part turns out to be the Laplace--Beltrami operator of the Perlick metric of type~I with $\beta=1$.  This solves in a natural way the ordering problem of the quantization
\begin{gather}\label{lb}
\hat{H}_{\rm LB3d} = -\frac{1}{2\sqrt{g}}\partial_i (g^{i,j}\sqrt{g} \partial_j)-  \mu\sqrt{\frac{1}{r^2} +k^2} -\frac{k^2}{2}, \qquad g= \det  g_{i,j}.
\end{gather}
We underline also the presence of a (maximal) degeneracy in the spectrum, a property which is characteristic of a M.S.\ quantum system
\begin{gather*}
\hat{H}_{\rm LB3d} \tilde{\psi}_{n,l+1}(r')Y_{l,m}(\theta,\phi) = \hat{H}_{\rm LB3d} \tilde{\psi}_{n',l'+1}(r')Y_{l,m}(\theta,\phi),\\
 n' \neq n,\qquad   l'\neq l, \qquad n'+l' = n+l .
\end{gather*}

\section{The conformal Perlick problem}\label{section4}

Let us rewrite the general Perlick I metric in a conformal way performing a point transformation:
\begin{gather}
ds^2 = \frac{dr^2}{\beta^2 (1+k^2r^2)} + r^2 d\Omega^2,  \qquad r \quad \Rightarrow  \quad \frac{2}{r^{-\beta}-k^2r^{\beta}} \label{pointct}\\
\Rightarrow  \quad ds^2=\frac{4}{r^2(r^{-\beta}-k^2r^{\beta})^2}\big(dr^2+r^2d\Omega^2\big).\nonumber
\end{gather}
The classical Hamiltonian  related to this new system of coordinates turns out to be:
\begin{gather*}
H = \frac{r^2(r^{-\beta}-k^2r^{\beta})^2}{8}\left(P_r^2  + \frac{L^2}{r^2} \right).
\end{gather*}
Seen from this perspective,  the physical meaning of this classical system appears to be twofold: in fact we could regard it  as describing either a particle with a constant mass in a curved space or   a particle with a non-constant mass $m =\frac{4}{r^2(r^{-\beta}-k^2r^{\beta})^2} $in a f\/lat space.

In Section~\ref{section3} we have investigated the hyperbolic Kepler system, namely the particular case of the Perlick~I systems with $\beta = 1$. Therefore in this case the transformation \eqref{pointct} is:
\begin{gather*}
r' = \frac{2\tilde{r}}{1-k^2\tilde{r}^2}.
\end{gather*}
In this new coordinate system  the Hamiltonian \eqref{lb} turns out to be:
\begin{gather*}
\hat{H}_{\rm LB} = -\frac{1}{8}\big(1-k^2\tilde{r}^2\big)^2 \left(\partial_{\tilde{r}}^2 + \frac{2k^2\tilde{r}}{1-k^2\tilde{r}^2}\partial_{\tilde{r}} +\frac{2}{\tilde{r}}\partial_{\tilde{r}} - \frac{q(q-1)}{\tilde{r}^2}\right) - \mu \left(\frac{1}{2\tilde{r}} + \frac{k^2\tilde{r}}{2}\right) -\frac{k^2}{2}.
\end{gather*}
 The above Hamiltonian is of course the LB quantization of the generalized Kepler in the new coordinate system. Now the question is: it is possible to get a ``physical'', though  ``nongeometrical'' quantization of this Perlick~I system without loosing the M.S.\ property?
The answer to this question is af\/f\/irmative, and indeed in the following we will introduce another M.S.\ quantization of this problem.

To this end, let us f\/irst  rewrite the scalar product in this new variable:
\begin{gather*}
\langle \psi_{n,l}|\psi_{n',l'}\rangle =\int_0^\infty \frac{{r'}^2}{\sqrt{1+k^2{r'}^2}} \psi_{n,l}^*(r') \psi_{n',l'}(r') dr'
\quad \Rightarrow \quad \int_D \frac{8\tilde{r}^2}{(1-k^2\tilde{r}^2)^3} \psi_{n,l}^*(\tilde{r}) \psi_{n',l'}(\tilde{r}) d\tilde{r}
\end{gather*}
with two possible domains:
\begin{gather*}
D = \tilde{r} \in  \left[0,\frac{1}{k}\right) ; \left(\frac{1}{k} , \infty\right).
\end{gather*}
Following the strategy of the previous  section we get:
\begin{gather*}
\epsilon_{n+l} = \int_D  \frac{8{\tilde{r}}^2}{(1-k^2\tilde{r}^2)^2} \frac{\psi_{n,l}^*(\tilde{r})}{\sqrt{1-k^2\tilde{r}^2}}  \left(  \frac{1}{\sqrt{1-k^2\tilde{r}^2}}  \hat{H}_{\rm LB}  \sqrt{1-k^2\tilde{r}^2}  \right)     \frac{\psi_{n,l}(\tilde{r})}{\sqrt{1-k^2\tilde{r}^2}} d\tilde{r}
\end{gather*}
we def\/ine now the new Hamiltonian:
\begin{gather}\label{hvm}
\hat{H}_{\rm vm} = \frac{1}{\sqrt{1-k^2\tilde{r}^2}}  \hat{H}_{\rm LB}  \sqrt{1-k^2\tilde{r}^2},\\
 \hat{H}_{\rm vm} = -\frac{1}{8}\big(1-k^2\tilde{r}^2\big)^2 \left(\partial_{\tilde{r}}^2  +\frac{2}{\tilde{r}}\partial_{\tilde{r}} - \frac{q(q-1)}{\tilde{r}^2}\right) - \mu \left(\frac{1}{2\tilde{r}} + \frac{k^2\tilde{r}}{2  }\right) -\frac{k^2}{8}.\nonumber
\end{gather}
We recognize easily the dif\/ferential part of this operator as the radial part of the Laplacian operator;
this is the so called Schr\"odinger quantization~\cite{iwaykatayama} which amounts to solve the order ambiguity in the quantization of the classical system putting the conformal factor in front of the dif\/ferential part, in other words the quantizations~\eqref{vonr} with parameters $a=1$, $b=0$.
Therefore in this case the Laplace--Beltrami quantization and the Schr\"odinger quantization  turn out to be linked through a similarity transformation, up to a constant shift $\frac{k^2}{8}$ proportional to the scalar curvature of this system. The set of eigenfunctions and eigenvectors are linked to each other, through easy algebraic manipulations.
In Section~\ref{section3} we showed the way to get by an algebraic computation the whole set of eigenfunctions of this Hamiltonian operator.

Before closing this section  we wish to summarize all the algebraic manipulations we have carried out, and to display  the general eigenfunctions written in a smarter way in term of Jacobi orthogonal polynomials $P_{(n,\alpha,\beta)}(x)$
\begin{gather*}
 \hat{H} = -\frac{1}{8}\big(1-k^2r^2\big)^2 \left(\partial_{r}^2  +\frac{2}{r}\partial_{r} - \frac{l(l+1)}{r^2}\right) - \mu
 \left(\frac{1}{2r} + \frac{k^2r}{2 }\right),\\
\psi_{n,l}(r) =  \frac{e^{-\frac{2 \mu \tanh^{-1} (kr)}{k(n+l+1)}}r^{n+l}}{(1-k^2r^2)^{n+l+\frac{1}{2}}} P \left(n,\frac{\mu- k(n+l+1)^2}{k(n+l+1)},-\frac{\mu-k(n+l+1)^2}{k(n+l+1)}\right) \left( \frac{1+k^2r^2}{2kr}\right),
\\
\hat{H}\psi_{n,l}(r) = -\left(\frac{\mu^2}{2(n+l+1)^2}+\frac{k^2(n+l+1)^2}{2} -\frac{k^2}{8}\right)\psi_{n,l}(r).
\end{gather*}

\begin{remark}
Laplace--Beltrami vs Schr\"odinger quantization
\end{remark}

The  above link between the Laplace--Beltrami and the Schr\"odinger quantization  is far from being a lucky occurrence.
Let us introduce a general conformal metric:
\begin{gather*}
ds^2  =  f(x_1,x_2,x_3)\big( dx_1^2+dx_2^2+dx_3^2\big).
\end{gather*}
 The scalar curvature of this conformal metric turns out to be:
\begin{gather*}
R(x_1,x_2,x_3) = \sum_i \frac{3{f_{x_i}}^2 - 4ff_{x_i x_i}}{2f^3}.
\end{gather*}
The classical Hamiltonian of a free particle in such space is:
\begin{gather*}
T=  \frac{1}{2 f(x_1,x_2,x_3)}\big(P_{x_1}^2 + P_{x_2}^2 + P_{x_3}^2\big).
\end{gather*}
Computing the Laplace--Beltrami quantization we get:
\begin{gather*}
\hat{T}_{\rm LB} = -\frac{1}{2\sqrt{g}}\partial_i\big(\sqrt{g}g^{i,j}\partial_j\big) =   -\frac{1}{2f} \left(\nabla^2 + \frac{ (\vec \nabla f) \cdot \vec \nabla}{f} \right),
\end{gather*}
 while the Schr\"odinger quantization leads to:
\begin{gather*}
\hat{T}_{\rm vm} = -\frac{1}{2f}\big(\nabla^2\big).
\end{gather*}
Now by a direct computation it is possible to verify that the two quantum operators are related by a similarity transformation, apart for an additive function which bears  a deep geometrical meaning, inasmuch as it represents the scalar curvature of the space:
\begin{gather*}
\frac{1}{f^{\frac{1}{4}}} \hat{T}_{\rm vm} f^{\frac{1}{4}} = \hat{T}_{\rm LB} + \frac{1}{16} R(x_1,x_2,x_3).
\end{gather*}
Therefore the spectrum of the two dif\/ferent quantizations coincide up to a constant shift only in the case of constant curvature: otherwise the scalar curvature term becomes a radial potential term and cannot any more be considered as a shift in the eigenvalues.

\section{Non-constant curvature quantum Perlick systems}\label{section5}

The machinery showed up to now plays a fundamental role in the quantization of the general Perlick problem~I, namely the metric~\eqref{perl1}  with $\beta \neq 1$. The general case $\beta \neq 1 $ is, geometrically speaking, a highly nontrivial space because the scalar curvature is no more constant as it was in the generalized Kepler:
\begin{gather*}
R(r) = -\frac{1}{2} \big(\big(\beta^2-1\big)\big(k^4r^{2\beta}+r^{-2\beta}\big)+2k^2\big(1+5\beta^2\big)\big).
\end{gather*}
In  the previous sections we performed the quantization following two dif\/ferent schemes, but since the scalar curvature turns out to be position-dependent the quantizations are no more similarity-equivalent  and we need to make a choice. The most suitable choice, which keeps the solvability property, turns out to be the quantization \eqref{vonr} with parameters $a=1$, $b=0$:
\begin{gather*}
\hat{H}_{vm} = - \frac{r^2(r^{-\beta}+k^2r^{\beta})^2}{8\beta^2}
\left(\partial_r^2 + \frac{2}{r}\partial_r - \frac{\hat{L}^2}{r^2}\right) -\frac{\mu}{2}\big(r^{-\beta} +k^2  r^{\beta}\big).
\end{gather*}
Performing the standard separation of variables:
\begin{gather*}
\psi_{nlm} (r,\theta_1 ,\theta_2)
= \phi(r)_{n,l}Y_{l,m}(\theta_1, \theta_2), \qquad \hat{L}^2 Y_{l,m}(\theta_1, \theta_2) = l(l+1) Y_{l,m}(\theta_1, \theta_2),
\end{gather*}
the Schr\"odinger problem is reduced to:
\begin{gather*}
 \left( - \frac{r^2(r^{-\beta}+k^2 r^{\beta})^2}{8\beta^2} \left(\partial_r^2 + \frac{2}{r}\partial_r - \frac{l(l+1)}{r^2}\right) -\frac{\mu}{2}\big(r^{-\beta} +k^2 r^{\beta}\big)\right) \phi_{n,l}(r) = E(n,l) \phi_{n,l}(r).
\end{gather*}
Now let us apply some algebraic manipulations already used in the past sections:
\begin{gather*}
r = {r'}^a, \qquad a = \frac{1}{\beta},
\\
\hat{H}_{\rm vm} = -\frac{(1-k^2 {r'}^2)^2}{8}\left(\partial_{r'}^2 +  \frac{1+a}{r'}\partial_{r'} + \frac{a^2 l (l+1)}{{r'}^2} \right)  -\frac{\mu}{2}\left(\frac{1}{r'} + k^2 r'\right),
\\
\hat{H}' = {r'}^{\frac{a-1}{2}} \hat{H}_{\rm vm} {r'}^{\frac{1-a}{2}} = -\frac{(1-k^2 {r'}^2)^2}{8}\!\left(\partial_{r'}^2 +  \frac{2}{r'} \partial_{r'} - \frac{a^2 l (l+1)\!-\!\frac{1-a^2}{4}  }{{r'}^2} \right)  -\frac{\mu}{2}\left(\frac{1}{r'} + k^2 r'\right).
\end{gather*}
This turns out to be exactly the Hamiltonian operator \eqref{hvm} if we set: $q = a l + \frac{a+1}{2}$.
Performing this substitution we found the new spectrum:
\begin{gather*}
E(n,l) = \frac{-\mu^2}{2(n+al+\frac{a+1}{2})^2} - \frac{k^2(n+al+\frac{a+1}{2})^2}{2} + \frac{k^2}{8}.
\end{gather*}
Let us consider $a$ as a rational number, i.e.\  $a = \frac{m_1}{m_2}$, $ m_1,m_2 \in {\mathbb{N}}$.
The two dif\/ferent wave functions $\psi_{n,l}$, $\psi_{n',l'}$ with $n' = n-s m_2$, $l'= l+sm_2$ $(s\in {\mathbb{N}} )$ are associated with the same energy eigenvalue, and therefore the degeneracy is still present in the general case $\beta \neq 1$ (but rational).

 Now it is crucial to underline that, since~$n$, $l$ are natural numbers,  the degeneracy of the spectrum can be preserved only if $a$ is a rational number too; but this is exactly the restriction requested also in the classical case for the closed orbit condition.

 Of course it is still possible to translate these results to the Laplace--Beltrami Hamiltonian by a suitable similarity transformation. Clearly,  in the LB case we would f\/ind a slightly dif\/ferent Hamiltonian with a potential correction term proportional to the scalar curvature:
\begin{gather*}
\frac{1}{f} \hat{H}_{\rm vm} f = \hat{H}_{\rm LB} + cR(r).
\end{gather*}

\section{Conclusions}

In the present paper, the f\/irst of a series on the radial quantum M.S., we have completely solved the f\/irst family of the quantum Perlick's systems with  parameter $k^2 >0$, giving contextually a~few insights about the ordering problem. The case with $k^2 < 0$ is also doable  and pretty much  similar
to the one investigated in the present paper;  the main dif\/ference consists just in the set of bounded states that turns out to be inf\/inite.

All the problems related to this family turn out to be solvable by  the factorization technique (SUSY quantum mechanics). We have also provided the general solutions of the eigenfunctions in terms of the Jacobi orthogonal polynomials.

The property of solvability and degeneracy in the spectrum holds indeed  for  the second family as well, providing therefore a complete classif\/ications of all the radial quantum M.S.\ Hamiltonians.

The results about the second family will be presented in a forthcoming paper.

\subsection*{Acknowledgments}
We wish to thank our colleagues and friends A.~Ballesteros, A.~Enciso and F.J.~Herranz for illuminating discussions  and crucial suggestions about the content of this paper.
The results reported here have been obtained in the framework of the INFN-MICINN collaboration 2010, and the related research activity has been partially supported by the Italian MIUR, through the PRIN 2008 research project n.20082K9KXZ/005.

\pdfbookmark[1]{References}{ref}
\LastPageEnding

\end{document}